\documentclass[
prd,preprint,aps,endfloats,tightenlines,
showpacs,superscriptaddress,nofootinbib,
amssymb
]{revtex4}
\usepackage{graphicx}
\begin{document}
\setlength{\baselineskip}{16pt}
%
%---------------------------------------------------------------------
%
\title{
Calculation of $K \to \pi\pi$ decay amplitudes
with an improved Wilson fermion action
in a nonzero momentum frame
in lattice QCD
}
%
%---------------------------------------------------------------------------
%
\author{ N.~Ishizuka }
\affiliation{
Center for Computational Sciences,
University of Tsukuba,
Tsukuba, Ibaraki 305-8577, Japan
}
\affiliation{
Graduate School of Pure and Applied Sciences,
University of Tsukuba,
Tsukuba, Ibaraki 305-8571, Japan
}
%--------------------------------------------
%
\author{ K.-I.~Ishikawa }
\affiliation{
Graduate School of Science, Hiroshima University,
Higashi-Hiroshima, Hiroshima 739-8526, Japan
}
%--------------------------------------------
%
\author{ A.~Ukawa }
\affiliation{
RIKEN Advanced Institute for Computational Science, 
Kobe, Hyogo 650-0047, Japan
}
\altaffiliation{
Renamed as RIKEN Center for Computational Science in April 2018
}
%--------
\affiliation{
Center for World Premier International Research Center Initiative (WPI), 
Japan Society for the Promotion of Science, 
Tokyo 102-0083, Japan
}
\altaffiliation{Address after April 2018}

%--------------------------------------------
%
\author{ T.~Yoshi\'{e} }
\affiliation{
Center for Computational Sciences,
University of Tsukuba,
Tsukuba, Ibaraki 305-8577, Japan
}
%-------
\affiliation{
Graduate School of Pure and Applied Sciences,
University of Tsukuba,
Tsukuba, Ibaraki 305-8571, Japan
}
%
%---------------------------------------------------------------------------
%
\date{ \today }
%%-- \date{ {\bf KPIPI\_p\_v01 : 20180514 : } \today }
%%-- \date{ {\bf KPIPI\_p\_v02 : 20180829 : } \today }
%%-- \date{ {\bf KPIPI\_p\_v03 : 20180831 : } \today }
%%-- \date{ {\bf KPIPI\_p\_v04 : 20180904 : } \today }
%%-- \date{ {\bf KPIPI\_p\_v05 : 20180904 : } \today }
%
%---------------------------------------------------------------------------
%
\begin{abstract}
We present our result for
the $K\to\pi\pi$ decay amplitudes
for both the $\Delta I=1/2$ and $3/2$ processes
with the improved Wilson fermion action.
In order to realize the physical kinematics,
where the pions in the final state have finite momenta,
we consider the decay process
$K({\bf p}) \to \pi({\bf p}) + \pi({\bf 0})$
in the nonzero momentum frame
with momentum ${\bf p}=(0,0,2\pi/L)$ on the lattice.
Our calculations are carried out
with $N_f=2+1$ gauge configurations
generated with the Iwasaki gauge action and
nonperturbatively $O(a)$-improved Wilson fermion action
at $a=0.091\,{\rm fm}$ ($1/a=2.176\,{\rm GeV}$),
$m_\pi=260\,{\rm MeV}$, and $m_K=570\,{\rm MeV}$
on a $48^3\times 64$ ($La=4.4\,{\rm fm}$) lattice.
For these parameters the energy of the $K$ meson
is set at that of two-pion in the final state.
We obtain
${\rm Re}A_2 = 2.431(19) \times10^{ -8}\,{\rm GeV}$,
${\rm Re}A_0 = 51(28)    \times10^{ -8}\,{\rm GeV}$, and 
$\epsilon'/\epsilon = 1.9(5.7) \times10^{-3}$ 
for a matching scale $q^* =1/a$ 
where the errors are statistical.  
The dependence on the matching scale $q^*$ of these values is weak.
The systematic error arising from
the renormalization factors is expected to be around 
$1.3\%$ for ${\rm Re}A_2$ and $11 \%$ for ${\rm Re}A_0$.
Prospects toward calculations with the physical quark mass are discussed.
\end{abstract}
\maketitle
%
%======================================================================
% @ Introduction
%
\section{ Introduction }
\label{Sec:Introduction}
Calculation of the $K\to\pi\pi$ decay amplitudes
is very important to quantitatively understand the $\Delta I=1/2$ rule
of the $K$ meson decay and
to theoretically predict the direct $CP$ violation parameter
($\epsilon'/\epsilon$) from the standard model.
A direct lattice QCD calculation of the decay amplitudes
for the $\Delta I=3/2$ process has been attempted for a long time.
The RBC-UKQCD Collaboration presented
the results at the physical quark mass in Ref.~\cite{A2:RBC-UKQCD_phys},
and those in the continuum limit in Ref.~\cite{A2:RBC-UKQCD_phys_cont},
which shows good agreement with experiment. 

A direct calculation of the decay amplitudes
for the $\Delta I=1/2$ process has been unsuccessful for a long time,
due to large statistical fluctuations coming from the disconnected diagrams.
A first direct calculation was reported by the RBC-UKQCD Collaboration
in Ref.~\cite{A0:RBC-UKQCD_400} at the pion mass $m_\pi=422\,{\rm MeV}$.
They also presented a result at a smaller quark mass ($m_\pi=330\,{\rm MeV}$)
at Lattice 2011~\cite{A0:RBC-UKQCD_300}.
They used the domain wall fermion action in their calculations
which preserves the chiral symmetry on the lattice.

Expanding on the earlier works
by Bernard {\it et al.}~\cite{CPS:Bernard} 
and by Donini {\it et al.}~\cite{CPS:Donini},
we showed in Ref.~\cite{A0:Our} 
that mixings with four-fermion operators with wrong chirality
are absent even for the Wilson fermion action
for the parity odd process 
in both the $\Delta I=3/2$ and $1/2$ channels due to $CPS$ symmetry.
Therefore, after subtraction of an effect from the lower dimensional operator,
a direct calculation of the decay amplitudes with the Wilson fermion action
is possible without complications from operators with wrong chirality,
as for the case with chirally symmetric lattice actions.
A potential advantage with the Wilson fermion action
over chirally symmetric lattice actions such as the domain wall action
is that the computational cost is generally smaller.
Hence, with the same amount of computational resources,
a statistical improvement may be expected with the lattice calculation
of the decay amplitudes, 
albeit this point has to be verified by actual calculations.
We reported a first result with the Wilson fermion action
at the pion mass $m_\pi = 280\,{\rm MeV}$ in Ref.~\cite{A0:Our}.

In both the RBC-UKQCD and our calculations
of the $\Delta I=1/2$ process referred above,
the kinematics was a $K$ meson at rest decaying to two zero momentum pions
at an unphysical quark mass satisfying $m_K \sim 2 m_\pi$.
The RBC-UKQCD Collaboration reported 
a first direct calculation in the physical kinematics,
where the pions in the final state have finite momenta,
at the physical quark mass in Ref.~\cite{A0:RBC-UKQCD_phys}.
They utilized the $G$-parity boundary condition in the spacial directions,
to extract the decay amplitude
from the $K\to\pi\pi$ time correlation function 
for the ground state of the two-pion
in the physical kinematics.

In the present work,
we consider the $K$ meson decay in the physical kinematics
by the nonzero momentum frame imposing the periodic boundary condition
in the spacial directions.
We calculate the decay amplitudes for the decay process
$K({\bf p}) \to \pi({\bf p}) + \pi({\bf 0})$
with momentum ${\bf p}=(0,0,2\pi/L)$ on the lattice.
We set the light quark mass
so that the energy of the initial $K$ meson
equals to that of the final two-pion state.
An advantage of using nonzero momentum 
in the periodic boundary condition over the $G$-parity boundary conditions 
is that gauge configurations,
generated previously with the normal boundary condition
for the other study and published on some databases,
can be used.

Our calculations have been carried out on 
the HA-PACS and the COMA at University of Tsukuba,
the K computer at the RIKEN Advanced Institute for Computational Science,
and the Oakforest-PACS 
at the Joint Center for Advanced High Performance Computing.
This paper is organized as follows.
In Sec.~\ref{Sec: Method}
we describe the method of calculation used in the present work.
The simulation parameters are also given.
We present our results in Sec.~\ref{Sec: Results}.
In Sec.~\ref{Sec: Conclusions}
conclusions of the present work are given 
and prospects toward calculations
with the physical quark mass are discussed.
%
%======================================================================
% @ Method
%
\section{ Method }
\label{Sec: Method}
%
%----------------------------------------------------------------------
% @ operator 
\subsection{ $\Delta S=1$ weak operators }
\label{Sec: Delta S=1 operators}
In this section the renormalization of the weak operators 
for the Wilson fermion action is reviewed.
The details have been shown in Ref.~\cite{A0:Our}.
The effective Hamiltonian of the $K\to\pi\pi$ decay
can be written as~\cite{Review:Buras}
\begin{equation}
  H = \frac{ G_F }{\sqrt{2}} \left( V_{us}^* V_{ud} \right)
      \sum_{i=1}^{10} \left( z_{i}(\mu) + \tau y_{i}(\mu) \right) Q_{i}(\mu)
\ ,
\label{eq:weak_Hamlitonian}
\end{equation}
with $\tau = - \left( V_{ts}^* V_{td} \right) 
             / \left( V_{us}^* V_{ud} \right)$, and
$z_{i}(\mu)$ and $y_{i}(\mu)$ ($i=1,2,\cdots 10$)
are the coefficient functions at renormalization scale $\mu$.
Here we consider the case $\mu \le m_{c}$,
where three light quarks, up, down and strange,
are the active quarks in the theory.
The ten operators $Q_{i}(\mu)$ ($i=1,2,\cdots 10$)
denote the $\Delta S=1$ four-fermion operators renormalized at $\mu$. 
The $K\to\pi\pi$ decay amplitude is obtained by
$A_I = \langle K | H | \pi\pi; I\rangle$ 
for the two-pion state with the isospin $I$.

The $K\to\pi\pi$ matrix elements $\langle K |Q_i|\pi\pi;I\rangle$
calculated on the lattice 
are converted to those in the continuum
by the renormalization factor for the operators $Q_i$. 
In the Wilson fermion action,
the chiral symmetry is broken explicitly. 
Hence mixings among different chiral multiplets are in general allowed,
and new operators arise through radiative corrections.
However, such a problem is absent for the parity odd part
of the operators of $Q_i$ due to the $CPS$ symmetry.

The mixing to lower dimensional operators is allowed.
From the $CPS$ symmetry and the equation of motion of the quark,
there is only one operator with the dimension less than 6,
which is
\begin{equation}
   Q_P = ( m_d - m_s ) \cdot \bar{s} \gamma_5 d
\ .
\label{eq:Q_P}
\end{equation}
This operator also appears in the continuum,
but does not yield nonvanishing contributions 
to the physical decay amplitudes,
since it is a total derivative operator.
However, this is not valid for the Wilson fermion action
due to chiral symmetry breaking,
and the operator (\ref{eq:Q_P}) does give nonzero unphysical contributions
to the amplitudes.
This contribution should be subtracted nonperturbatively,
because the mixing coefficient includes
a power divergence due to the lattice cutoff growing as $1/a^2$.
We can subtract it as
\begin{equation}
  \overline{Q}_i = Q_i - \beta_i Q_P 
\ , 
\label{eq:sub_Q}
\end{equation}
by imposing the following condition~\cite{sub:Dawson,sub:Testa},
\begin{equation}
    \langle 0 | \, \overline{Q}_i             \,| K \rangle
 =  \langle 0 | \, Q_i -  \beta_i \cdot Q_P   \,| K \rangle = 0
\ ,
\label{eq:sub_Q_rel}
\end{equation}
for each operator $Q_i$.
The matrix of the renormalization factor
of the subtracted operators $\overline{Q}_i$
has the same structure as in the case in which chiral symmetry is preserved. 
%
%---------------------------------------------------------------------
% @ Simulation parameters
\subsection{ Simulation parameters }
Our calculations are carried out
with $N_f=2+1$ gauge configurations
generated with the Iwasaki gauge action 
and nonperturbatively $O(a)$-improved Wilson fermion action
on a $48^3\times 64$ at $\beta=1.9$. 
The hopping parameters are 
$\kappa_{ud}=0.137712$ for the up and the down quark, and
$\kappa_{s }=0.136400$ for the strange quark.
We use the same algorithm for a generation of the gauge configurations, 
and the same value of $\beta$ and the improved factor $C_{SW} = 1.7150$
as in Ref.~\cite{conf:PACS-CS}.

We measure hadron Green's functions
and the $K$ meson decay amplitudes at every $10$ trajectories.
The total length of the run is $9800$ trajectories
and the total number of gauge configurations employed
for the measurement is $980$.
We estimate statistical errors by the jackknife method
with bins of $20$ configurations ($200$ trajectory).
The parameters determined from the spectrum analysis
are $a=0.091\,{\rm fm}$ ($1/a=2.176\,{\rm GeV}$) for lattice spacing,
$La=4.37\,{\rm fm}$ for spatial lattice size,
and 
\begin{eqnarray}
  &&  m_\pi  = 0.11794(33)\  (\, 256.63(72)\,{\rm MeV}\, )  \ , \cr 
  &&  m_K    = 0.26275(20)\  (\, 571.74(43)\,{\rm MeV}\, )  \ ,
\end{eqnarray}
for the pion and the $K$ meson mass.
%%--
%%--  1/a = 2.176 GeV
%%--   1.179351e-01 * 2.176 * 1000 = 256.6267776
%%--       3.32D-04 * 2.176 * 1000 =   0.722432
%%--   2.627472e-01 * 2.176 * 1000 = 571.7379072
%%--       1.97e-04 * 2.176 * 1000 =   0.428672 

In the present work,
we consider the $K\to\pi\pi$ decay in the physical kinematics,
where the pions in the final state have finite momenta.
In order to realize this,
we calculate the decay amplitudes for the decay process
$K({\bf p}) \to \pi({\bf p}) + \pi({\bf 0})$
in the nonzero momentum frame
with momentum ${\bf p}=(0,0,2\pi/L)$ on the lattice.
We set the light quark mass 
so that the energy difference between the initial $K$ meson
and the final two-pion state vanishes.
On our gauge configurations
the energy difference is 
$\Delta E \equiv E_K - E_{\pi\pi}^I = -4.7(1.4)\,{\rm MeV}$ for $I=2$
                                  and $9.8(7.8)\,{\rm MeV}$ for $I=0$
as shown in the following section.
In the present work,
we assume that these mismatches of the energy
give only small effects to the decay amplitudes.
%%-
%%-  1/a = 2.176 GeV
%%-  ( 2.930304e-01 - 2.951910e-01 ) * 2.1760 * 1000 = -4.7014
%%-                     ( 6.23e-04 ) * 2.1760 * 1000 =  1.3556
%%-  ( 2.930304e-01 - 2.885337e-01 ) * 2.1760 * 1000 =  9.7848 
%%-                     ( 3.60e-03 ) * 2.1760 * 1000 =  7.8336
%%-
%
%---------------------------------------------------------------------
% @ Time correlation functions
\subsection{ Time correlation function for $K\to\pi\pi$ }
We extract the matrix element
$\langle K | \overline{Q}_i | \pi\pi ; I \rangle$
from the time correlation function for the $K\to\pi\pi$ process,
\begin{equation}
  G^{I}_{i}(t,t_1,t_2) = \frac{1}{T} \sum_{\delta=0}^{T-1}
       \langle 0 | \, 
           W_{K^0}(t_K+\delta) \,\, 
           \overline{Q}_{i}(t+\delta) \,\,
           W_{\pi\pi}^{I}(t_1+\delta, t_2+\delta) 
       \, | 0 \rangle
\ ,
\label{eq:G_KPIPI}
\end{equation}
where $\overline{Q}_i(t)$ is the subtracted weak operator
at the time slice $t$ defined by
\begin{equation}
  \overline{Q}_i(t) = \sum_{\bf x} \overline{Q}_i({\bf x},t)
\ ,
\end{equation}
with the subtracted operator $\overline{Q}_i({\bf x},t)$
at the space-time position $({\bf x},t)$
defined in (\ref{eq:sub_Q}).

The operator $W_{K^0}(t)$ in (\ref{eq:G_KPIPI})
is the wall source for the $K^0$ meson
with the momentum ${\bf p}=(0,0,-p)$ ($p=2\pi/L$)
at the time slice $t$,
\begin{equation}
 W_{K^0}(t) =  
   - \left[\    \overline{W}_{d}(-p,t) \gamma_5 W_{s}( 0,t)
              + \overline{W}_{d}( 0,t) \gamma_5 W_{s}(-p,t) 
     \ \right]/2
\ ,
\label{eq:wsource_K}
\end{equation}
where the wall source for the quark $q=u,d,s$ 
with the momentum ${\bf k}=(0,0,k)$
is given by
\begin{eqnarray}
&&           {W}_{q}(k,t) =
                  \sum_{\bf x} q ({\bf x}, t)\,
                        {\rm e}^{  i{\bf x}\cdot {\bf k}}\ , \\
&&  \overline{W}_{q}(k,t) = 
                  \sum_{\bf x} \bar{q}({\bf x}, t)\,
                        {\rm e}^{  i{\bf x}\cdot {\bf k}}\ .
\end{eqnarray}
We adopt $K^0 = - \bar{d} \gamma_5 s$ as the neutral $K$ meson operator,
so our correlation function has an extra minus from the usual convention.

The operator $W_{\pi\pi}^{I}(t_1,t_2)$ in (\ref{eq:G_KPIPI})
is the wall source
for the two-pion state with the isospin $I$
with the momentum ${\bf p}=(0,0,p)$ ($p=2\pi/L$), 
\begin{equation}
  W_{\pi\pi}^{I}(t_1,t_2) =
   \left[\  
      \tilde{W}_{\pi\pi}^{I}(k_1,k_2,t_1,t_2)
       + \left( t_1 \leftrightarrow t_2 \right)
       + \left( k_1 \leftrightarrow k_2 \right)
       + \left( t_1 \leftrightarrow t_2 ,
                k_1 \leftrightarrow k_2 \right)
     \ \right]/4
\ , 
\label{eq:W_pipiI}
\end{equation}
with $k_1=p=2\pi/L$ and $k_2=0$, 
where 
\begin{eqnarray}
&&  \tilde{W}_{\pi\pi}^{I=2}(k_1,k_2,t_1,t_2) =
       \Bigl(   W_{\pi^0}(k_1,t_1) W_{\pi^0}(k_2,t_2)
              + W_{\pi^+}(k_1,t_1) W_{\pi^-}(k_2,t_2)
       \Bigr)/\sqrt{3}
\ ,
\label{eq:W_pipi2}
\\
&&  \tilde{W}_{\pi\pi}^{I=0}(k_1,k_2,t_1,t_2) =
       \Bigl(           - W_{\pi^{0}}(k_1,t_1) W_{\pi^{0}}(k_2,t_2) /\sqrt{2}
               + \sqrt{2} W_{\pi^{+}}(k_1,t_1) W_{\pi^{-}}(k_2,t_2)
       \Bigr)/\sqrt{3}
\ . \qquad 
\label{eq:W_pipi0}
\end{eqnarray}
The operator $W_{\pi^{i}}(k,t)$ 
is the wall source for $\pi^i$ meson with the momentum ${\bf k}=(0,0,k)$ 
at the time slice $t$,
\begin{equation}
     W_{\pi^{i}}(k,t) = 
       \left[ \ \tilde{W}_{\pi^{i}}(k_1,k_2,t)
                 + \left( k_1 \leftrightarrow k_2 \right)
       \ \right]/2
\ , 
\label{eq:W_pit}
\end{equation}
with $k_1=k$ and $k_2=0$, 
where 
\begin{eqnarray}
&&    \tilde{W}_{\pi^{+}}(k_1,k_2,t) = 
             - \overline{W}_d(k_1,t) \gamma_5 W_u(k_2,t)
\ ,
\label{eq:W_pip}
\\
&&    \tilde{W}_{\pi^{0}}(k_1,k_2,t) =  
               \left(\, \overline{W}_u(k_1,t) \gamma_5 W_u(k_2,t)
                      - \overline{W}_d(k_1,t) \gamma_5 W_d(k_2,t) 
               \, \right)/\sqrt{2}
\ ,
\label{eq:W_pi0}
\\
&&    \tilde{W}_{\pi^{-}}(k_1,k_2,t) =   
              \overline{W}_u(k_1,t) \gamma_5 W_d(k_2,t)
\ .
\label{eq:W_pim}
\end{eqnarray}

The periodic boundary condition is imposed in all directions.
The summation over $\delta$,
where $T=64$ denotes the temporal size of the lattice,
is taken in (\ref{eq:G_KPIPI}) to improve the statistics.
The time slice of the $K$ meson is set at $t_K=29$.
The wall source of each pion is separated by four lattice unit 
according to $t_1=0$ and $t_2=4$ in (\ref{eq:G_KPIPI})
to improve the statistics
following the suggestion by Ref.~\cite{A0:RBC-UKQCD_phys}.
The gauge configurations are fixed to the Coulomb gauge
at the time slice of the wall source 
$t_K+\delta$, $t_1+\delta$ and $t_2+\delta$
for each $\delta$.

The mixing coefficient 
of the lower dimensional operator, $\beta_i$ in (\ref{eq:sub_Q}), 
is obtained from the following ratio of the time correlation function $K\to 0$,
\begin{equation}
  \beta_{i} =
    \sum_{\delta_1=0}^{T-1} \langle 0 | \, W^{0}_{K^0}(t_K+\delta_1) 
                                        \, Q_{i}(t+\delta_1) 
                        \, | 0 \rangle
    \Bigl/ \,
    \sum_{\delta_2=0}^{T-1} \langle 0 | \, W^{0}_{K^0}(t_K+\delta_2) 
                                        \, Q_P(t+\delta_2) 
                        \, | 0 \rangle
\ ,
\label{eq:calc_alpha}
\end{equation}
in the large $t_K-t$ region,
where $W^{0}_{K^0}(t)$ is the wall source for the $K^0$ meson 
with the zero momentum 
given from (\ref{eq:wsource_K}) by setting $p=0$, and 
$Q_i(t)=\sum_{\bf x} Q_i({\bf x},t)$ and
$Q_P(t)=\sum_{\bf x} Q_P({\bf x},t)$,
with the operator ${Q}_i({\bf x},t)$
and ${Q}_P({\bf x},t)$ defined in (\ref{eq:Q_P})
at the space-time position $({\bf x},t)$.

The quark contractions of the $K\to\pi\pi$ time correlation function 
$G_i^I(t,t_1,t_2)$ in (\ref{eq:G_KPIPI}) are given by
(57), (58) and (59) in Ref.~\cite{A0:Our}.
A subtraction of the contributions of the vacuum diagram,
$\langle 0 | W_{K^0}(t_K+\delta) \, \overline{Q}_i(t+\delta) | 0 \rangle 
 \langle 0 | W_{\pi\pi}^{I}(t_1+\delta,t_2+\delta) | 0 \rangle$,
is not necessary in the present case of the nonzero momentum frame.
Those of the $K\to 0$ time correlation function 
in (\ref{eq:calc_alpha}) are given by (60), (61) and (62) 
in Ref.~\cite{A0:Our}.

The quark loop at the weak operator,
the quark propagator starting from the position of the weak operator
and ending at the same position,
appears in the quark contractions
for the $K\to\pi\pi$ and the $K\to 0$ processes.
We calculate them by the same method 
as the previous work in Ref.~\cite{A0:Our},
where the stochastic methods 
with the hopping parameter expansion technique (HPE)
and the truncated solver method (TSM)
proposed in Ref.~\cite{HPE_RSM:Bali} are used.
As found in the previous work, 
we find that 
the first term of TSM in (77) in Ref.~\cite{A0:Our}
is negligible for all time correlation functions also in the present work,
when the stopping condition is set at  
$R < 1.2 \times 10^{-6}$.
Thus we adopt $N_R=0$ and $N_T=10$ with this stopping condition 
in the calculation of the quark loop.
%
%---------------------------------------------------------------------
% @ pipi->pipi
\subsection{ Time correlation function for $\pi\pi\to\pi\pi$ }
We calculate two types of time correlation functions 
for the $\pi\pi\to\pi\pi$ process
to obtain the normalization factors
which are  needed to extract the matrix elements
$\langle K | \overline{Q}_i | \pi\pi ; I \rangle$
from the time correlation function
$G^{I}_{i}(t,t_1,t_2)$ in (\ref{eq:G_KPIPI}).
These are point-wall and wall-wall time correlation functions,
which are defined by
\begin{eqnarray}
&&  G_{PW}^{I}(t,t_1,t_2) =
          \frac{1}{T} \sum_{\delta=0}^{T-1}
          \langle 0 | \, (\pi\pi)^{I}(t+\delta)\,\,
                         W_{\pi\pi}^{I}(t_1+\delta, t_2+\delta) 
          \, | 0 \rangle
\ ,
\label{eq:G_PIPI_PW}
\\
&&  G_{WW}^{I}(t,t_1,t_2) = 
         \frac{1}{T} \sum_{\delta=0}^{T-1}
         \langle 0 | \, W_{\pi\pi}^{I}(t  +\delta, t+[t_2-t_1]+\delta ) \,\,
                        W_{\pi\pi}^{I}(t_1+\delta, t_2+\delta) 
         \, | 0 \rangle
\ ,
\label{eq:G_PIPI_WW}
\end{eqnarray}
The operator $(\pi\pi)^{I}(t)$ is the operator 
for the two-pion with the isospin $I$
with the momentum ${\bf p}=(0,0,-p)$ ($p=2\pi/L$), 
\begin{equation}
   (\pi\pi)^{I}(t) =
         \left[\, 
             \overline{(\pi\pi)}^{\ I}(k_1,k_2,t)
           + \left( k_1 \leftrightarrow k_2 \right)
         \,\right] / 2
\ , 
\label{eq:psource_pipit}
\end{equation}
with $k_1=-p$ and $k_2=0$, 
where 
\begin{eqnarray}
&&  \overline{(\pi\pi)}^{\ I=2}(k_1,k_2,t) =
         \Bigl(   \pi^{0}(k_1,t) \pi^{0}(k_2,t)
                + \pi^{+}(k_1,t) \pi^{-}(k_2,t)
         \Bigr)/\sqrt{3}
\ ,
\label{eq:psource_pipi2}
\\
&&  \overline{(\pi\pi)}^{\ I=0}(k_1,k_2,t) =
         \Bigl( -          \pi^{0}(k_1,t) \pi^{0}(k_2,t) /\sqrt{2}
                + \sqrt{2} \pi^{+}(k_1,t) \pi^{-}(k_2,t)
         \Bigr)/\sqrt{3}
\ .
\label{eq:psource_pipi0}
\end{eqnarray}
The operator $\pi^{i}(k,t)$ is the operator 
for $\pi^{i}$ meson with the momentum 
${\bf k}=(0,0,k)$, defined by
\begin{eqnarray}
&&    \pi^{+}(k,t) =
         - \sum_{\bf x}\  
               \bar{d}({\bf x},t) \gamma_5 u({\bf x},t) 
               \cdot {\rm e}^{ i {\bf x}\cdot{\bf k} }
\ ,
\label{eq:P_pip}
\\
&&    \pi^{0}(k,t) =
          \sum_{\bf x}\ 
            \left(\,  \bar{u}({\bf x},t) \gamma_5 u({\bf x},t)
                    - \bar{d}({\bf x},t) \gamma_5 d({\bf x},t)
            \, \right)/\sqrt{2}
            \cdot {\rm e}^{ i {\bf x}\cdot{\bf k} }
\ ,
\label{eq:P_pi0}
\\
&&    \pi^{-}(k,t) = 
           \sum_{\bf x}\ 
               \bar{u}({\bf x},t) \gamma_5 d({\bf x},t)
               \cdot {\rm e}^{ i {\bf x}\cdot{\bf k} }
\ .
\label{eq:P_pim}
\end{eqnarray}
The operator $W_{\pi\pi}^{I}(t_1,t_2)$ ($I=0,2$) is defined by
(\ref{eq:W_pipi2}) and (\ref{eq:W_pipi0}).
In the present work, we fix $t_1=0$ and $t_2=4$ 
in (\ref{eq:G_PIPI_PW}) and (\ref{eq:G_PIPI_WW}) 
as in the $K\to\pi\pi$ process.

The quark contractions for the $\pi\pi\to\pi\pi$ time correlation function
are given by (89) and (90) in Ref.~\cite{A0:Our}.
Subtractions of the contributions of the vacuum diagrams,
$\langle 0 | (\pi\pi)^{I}(t+\delta) | 0 \rangle
 \langle 0 | W_{\pi\pi}^{I}(t_1+\delta,t_2+\delta) | 0 \rangle$
and
$\langle 0 | W_{\pi\pi}^{I}(t  +\delta,t + [t_2-t_1] +\delta) | 0 \rangle
 \langle 0 | W_{\pi\pi}^{I}(t_1+\delta,t_2+\delta) | 0 \rangle$,
are not necessary in the present case of the nonzero momentum frame.
%
%======================================================================
% @ Results
%
\section{ Results }
\label{Sec: Results}
%
%---------------------------------------------------------------------
% @ Time correlation function
\subsection{ Energy of two-pion state }
\label{Sec: Energy of two-pion state }
The time correlation functions for the $\pi\pi\to\pi\pi$ process 
$G_{PW}^{I}(t,t_1,t_2)$ in (\ref{eq:G_PIPI_PW}) and 
$G_{WW}^{I}(t,t_1,t_2)$ in (\ref{eq:G_PIPI_WW})
behave in the large time region as
\begin{eqnarray}
&&  
  G_{PW}^{I}(t,t_1,t_2) = 
      A^I \cdot f(t,t_\pi,E_{\pi\pi}^{I})
    + C^I \cdot f(t,t_\pi,S_\pi)
\ ,
\label{eq:fit_G_PIPI_PW}
\\
&&
  G_{WW}^{I}(t,t_1,t_2) = 
      A_{\pi\pi}^{I} \cdot f(t,t_\pi,E_{\pi\pi}^{I})
    + D^I            \cdot f(t,t_\pi,S_\pi)
\ ,
\label{eq:fit_G_PIPI_WW}
\end{eqnarray}
with $t_\pi = (t_1+t_2)/2$,
where 
\begin{equation}
   f(t,t_\pi,m) = {\rm e}^{ - m \cdot       | t - t_\pi |   } 
                + {\rm e}^{ - m \cdot ( T - | t - t_\pi | ) }
\ , 
\label{eq:func_f}  
\end{equation}
$E_{\pi\pi}^{I}$ is the energy of the two-pion state with the isospin $I$,
$S_\pi = \sqrt{m_\pi^2 + p^2 } - m_\pi$,
$A^I$ is a constant whose form is irrelevant,
and
\begin{equation}
   A_{\pi\pi}^I
     =    \langle 0 |\, W_{\pi\pi}^{I}(\Delta,-\Delta) \, | \pi\pi; I \rangle^2
        / \langle \pi\pi; I | \pi\pi; I \rangle
\ ,
\label{eq:A_PIPI}
\end{equation}
with $\Delta = ( t_2 - t_1 )/2$.
The terms with the constant $C^I$ and the $D^I$
in (\ref{eq:fit_G_PIPI_PW}) and (\ref{eq:fit_G_PIPI_WW})
arise from the process in which either of the two pions propagates 
in the opposite time direction to the other pion 
({\it i.e.,} around-the-world effect for the two-pion operator).

\begin{figure}[h]
\includegraphics[height=6.5cm]{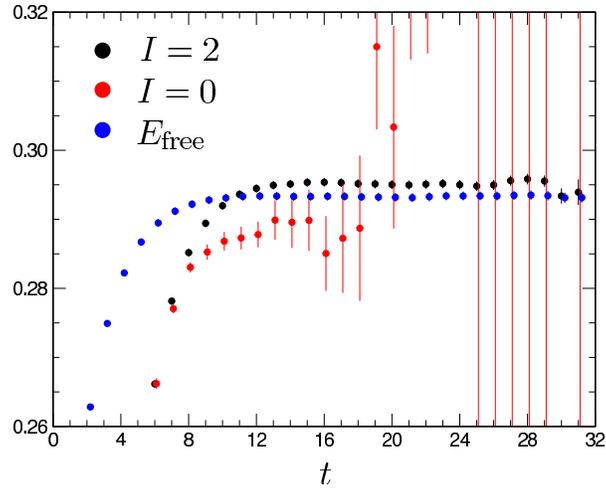}
\caption{
Effective energy of the time correlation function
$G_{PW}^{I}(t,t_1,t_2)$ for the $\pi\pi\to\pi\pi$
with the isospin $I=0$ and $I=2$.
The source operator for the two-pion is located at $(t_1,t_2)=(0,4)$.
A value in the non-interaction case
given by adding the effective energy of the pion 
with the momentum ${\bf p}=(0,0,2\pi/L)$ and 
that for the zero momentum, 
which corresponds to
$E_{\rm free}=\sqrt{m_\pi^2+p^2} + m_\pi$, 
is plotted by blue symbol for a comparison.
}
\label{fig:LM_PIPI02_LW}
%
%------------
\newpage
%------------
\end{figure}

The effective energy of the point-wall time correlation function 
$G_{PW}^{I}(t)$
is plotted in Fig.~\ref{fig:LM_PIPI02_LW},
where the effective energy $E_{eff}$ at $t$ is obtained by
\begin{eqnarray}
&&
     \frac{   f(t+4,t_\pi,S_\pi) \cdot G_{PW}^{I}(t+1,t_1,t_2)
            - f(t+1,t_\pi,S_\pi) \cdot G_{PW}^{I}(t+4,t_1,t_2) } 
          {   f(t+3,t_\pi,S_\pi) \cdot G_{PW}^{I}(t  ,t_1,t_2)
            - f(t  ,t_\pi,S_\pi) \cdot G_{PW}^{I}(t+3,t_1,t_2) } 
\cr \qquad 
&&
   = 
     \frac{   f(t+4,t_\pi,S_\pi) \cdot f(t+1,t_\pi,E_{eff})
            - f(t+1,t_\pi,S_\pi) \cdot f(t+4,t_\pi,E_{eff}) } 
          {   f(t+3,t_\pi,S_\pi) \cdot f(t  ,t_\pi,E_{eff})
            - f(t  ,t_\pi,S_\pi) \cdot f(t+3,t_\pi,E_{eff}) } 
\ ,
\label{eq:pipi_localmass_fit}
\end{eqnarray}
with determined value of $S_\pi = \sqrt{m_\pi^2 + p^2 } - m_\pi$ 
from the correlation function of the pion.
In the present work we set the source operator for the two-pion 
at $(t_1,t_2)=(0,4)$
as mentioned in the previous subsection.

We find plateaus in the time region $t \ge 13$ for both $I=0$ and $2$.
The value in the noninteracting two-pion case obtained by 
adding the effective energy of the pion 
with the momentum ${\bf p}=(0,0,2\pi/L)$ and 
that for the zero momentum,
which corresponds to $E_{\rm free}=\sqrt{m_\pi^2+p^2} + m_\pi$, 
is plotted for comparison.
We find that 
the two-pion energy for $I=2$ is larger than $E_{\rm free}$
signifying repulsive interaction of the two-pion system,
whereas that for $I=0$ is smaller showing attractive interaction.

In the extraction of
the matrix elements
$\langle K | \overline{Q}_i | \pi\pi ; I \rangle$
from the time correlation function
$G^{I}_{i}(t,t_1,t_2)$ in (\ref{eq:G_KPIPI}),
the values of $E_{\pi\pi}^{I}$ and $A_{\pi\pi}^I$ are needed.
Since the statistical error of the point-wall correlation function 
$G_{PW}^{I}(t,t_1,t_2)$ is smaller than
that for the wall-wall function $G_{WW}^{I}(t,t_1,t_2)$,
we first extract the energy $E_{\pi\pi}^{I}$ from $G_{PW}^{I}(t,t_1,t_2)$
with the determined value of $S_\pi = \sqrt{m_\pi^2 + p^2 } - m_\pi$ 
from the  correlation function of the pion.
We then extract the amplitude $A_{\pi\pi}^I$ from $G_{WW}^{I}(t,t_1,t_2)$
by fitting to (\ref{eq:fit_G_PIPI_WW})
with the determined value of $E_{\pi\pi}^{I}$ and $S_\pi$,
and regarding $A_{\pi\pi}^I$ and $D^I$ as unknown parameters.
The results for $E_{\pi\pi}^{I}$ and $A_{\pi\pi}^{I}$ are
\begin{eqnarray}
&&   E_{\pi\pi}^{I=2} = 0.29519(62)                \ , \qquad
     A_{\pi\pi}^{I=2} = 1.0734(67) \times 10^{21}  
\ ,
\label{eq:table_PIPI_E_2}
\\
&&   E_{\pi\pi}^{I=0} = 0.2885(36)                 \ \ , \qquad
     A_{\pi\pi}^{I=0} = 1.034(39)  \times 10^{21}  
\  ,
\label{eq:table_PIPI_E_0}
\end{eqnarray}
in the lattice unit,
where we adopt the fitting range
$t=[14,26]$ for $I=2$ and $t=[14,20]$ for $I=0$.
%%-
%%-    4 : LW_PP_2  :  14 26 :  2.951910D-01  6.23D-04 :  3.80D-05
%%-    5 : LW_PP_0  :  14 20 :  2.885337D-01  3.60D-03 :  2.06D-03
%%-   10 : WW_PP_2  :  14 26 :  1.073398D+21  6.71D+18 :  8.61D-04
%%-   11 : WW_PP_0  :  14 20 :  1.033500D+21  3.93D+19 :  7.44D-02
%-%
The energy difference between the initial $K$ meson and the final two-pion state
$\Delta E^I = E_K - E_{\pi\pi}^I$ ($E_K=\sqrt{m_K+p^2}$, $p=2\pi/L)$ are
\begin{eqnarray}
  && \Delta E^{I=2} =     -0.00216(62)\    (\, -4.7(1.4) \,{\rm MeV}\, ) \ ,  \\
  && \Delta E^{I=0} = \ \  0.0045(36) \ \  (\,  9.8(7.8) \,{\rm MeV}\, ) \ .
\end{eqnarray}
In the present work,
we assume that these violations of energy conservation
yield only small effects to the results for the $K\to\pi\pi$ decay amplitudes.
%%-
%%-  ( 2.930304e-01 - 2.951910e-01 ) * 2.1760 * 1000 = -4.7014
%%-                     ( 6.23e-04 ) * 2.1760 * 1000 =  1.3556
%%-  ( 2.930304e-01 - 2.885337e-01 ) * 2.1760 * 1000 =  9.7848 
%%-                     ( 3.60e-03 ) * 2.1760 * 1000 =  7.8336
%%-
%---------------------------------------------------------------------
% @ Matrix elements
\subsection{ $K\to\pi\pi$ matrix elements }
\label{Sec: Matrix elements}
In order to extract the $K\to\pi\pi$ matrix element,
we consider an effective matrix element $M^{I}_{i}(t)$,
which behaves as
$M^{I}_{i}(t) 
  = M^I_i \equiv \langle K |\, 
     \overline{Q}_i({\bf x},0)\, 
   | \pi\pi; I \rangle$
in the time region $(t_1,t_2) \ll t \ll t_K$.
It can be constructed from the time correlation function
$G^{I}_{i}(t,t_1,t_2)$ in (\ref{eq:G_KPIPI}) by
\begin{equation}
  M^{I}_{i}(t) = \frac{G^{I}_{i}(t,t_1,t_2)}
   {\sqrt{ A_K A_{\pi\pi}^{I} }}
   \cdot F^{I}
   \cdot {\rm e}^{ E_K (t_K-t) + E^{I}_{\pi\pi} (t-t_\pi) }
   \times (-1)
\ ,
\label{eq:effective_amp}
\end{equation}
where $t_\pi = ( t_1 + t_2 )/2$.
Here, the energy of the $K$ meson $E_K$, 
and the energy of the two-pion state $E^{I}_{\pi\pi}$ 
are fixed at the values
obtained from the time correlation function of the $K$ meson
and the $\pi\pi\to\pi\pi$.
The factor $(-1)$ comes from our convention of the $K^0$ operator
in (\ref{eq:wsource_K}).
The constant
$A_K = \langle 0 | W_K | K \rangle^2 / \langle K | K \rangle$
is estimated from the wall-wall propagator of the $K$ meson,
with the value 
$A_K=2.3544(48)\times 10^{10}$ 
in the lattice unit.
The constant $A_{\pi\pi}^{I}$ is defined by (\ref{eq:A_PIPI})
and its value is given 
by (\ref{eq:table_PIPI_E_2}) or (\ref{eq:table_PIPI_E_0}).

The constant $F^{I}$ in (\ref{eq:effective_amp}) 
is the Lellouch-L\"uscher factor~\cite{LL-factor:LL}
in the nonzero momentum frame
with the momentum $|{\bf p}|=p$ given 
in Refs.~\cite{LL-factor:KSS,LL-factor:CKY},
\begin{eqnarray}
  ( F^{I} )^2
     &=& \langle K | K \rangle \cdot
         \langle \pi\pi ; I | \pi\pi ; I \rangle / V^2
\cr
     &=& (4\pi)\left( \frac{ ( E^{I}_{\pi\pi} )^2 \, m_K }{ p_c^3 } \right)
               \left( p_c \frac{ \partial\delta^{I}(p_c) }{\partial p_c}
                    + q_c \frac{ \partial\phi(q_c) }{\partial q_c} \right)
 \ ,
\label{eq:F_LL}
\end{eqnarray}
where $V$ is the lattice volume $V=L^3$,
$\delta^{I}(p_c)$ is the two-pion scattering phase shift
for the isospin $I$
at the scattering momentum $p_c^2=( (E^{I}_{\pi\pi})^2 - p^2 )/4 - m_\pi^2$,
and the function $\phi(q_c)$ at $q_c = p_c \cdot L/(2\pi)$ is defined 
in Refs.~\cite{LL-factor:KSS,LL-factor:CKY}.
In the noninteracting two-pion case,
the factor takes the form
$(F^{I})^2 \equiv ( F|_{\rm free} )^2 = (m_\pi + E_\pi)^2 E_K V$,
where $E_\pi=\sqrt{m_\pi^2 + p^2}$ and $E_K=\sqrt{m_K^2 + p^2}$
($p=2\pi/L$).
For the interacting case, we need the value of the first derivative 
of the phase shift $\delta^I(p_c)$
with the momentum $p_c$ to calculate the factor.
However, the phase at only one momentum is calculated in the present work
and the derivative cannot be evaluated.
In the present work,
we estimate the factor with an approximation,
\begin{equation}
  \delta^I(p_c) =
           p_c \frac{ \partial \delta^I(p_c) }{ \partial p_c }
         + {\rm O}(p_c^3)
\ , 
\label{eq:appx_delta}
\end{equation}
neglecting the cubic term, 
leaving a precise estimation of the factor to study in the future. 
The calculated values of the Lellouch-L\"uscher factors are 
tabulated in Table~\ref{table:F_LL}.
We find that the main contribution to the factor comes from 
the term $q_c \cdot ( \partial \phi(q_c) / \partial q_c )$
while the term with the derivative of the phase shift is small
($2\%$ for the $I=2$ and $12\%$ for the $I=0$ case).
We consider from this 
that our approximation for the phase shift in (\ref{eq:appx_delta})
does not make a large systematic error.

\begin{figure}[h]
\includegraphics[width=8.2cm]{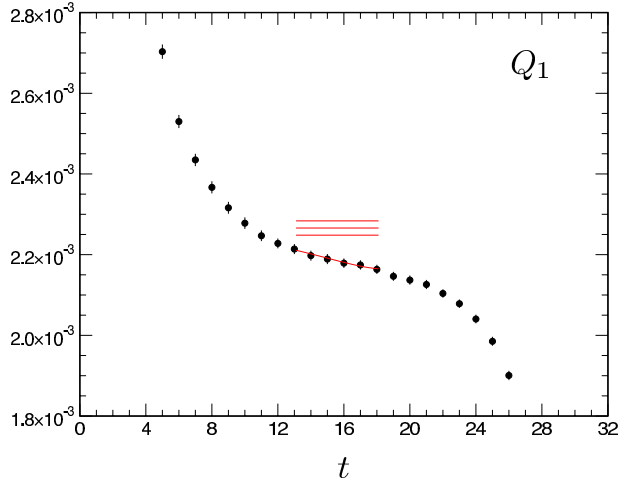}
\caption{
Effective matrix element
of $M_{i}^{I=2}(t)$ in (\ref{eq:effective_amp}) 
for the operator $Q_1$ in the lattice unit.
The operator for the two-pion is located at $(t_1,t_2)=(0,4)$
and the $K$ meson at $t_K=29$.
The operator $Q_i(t)$ runs over the whole time extent.
The result of the fitting to (\ref{eq:effective_amp_tdep})
in the time range $t=[13,18]$
is plotted by a curved line 
and the result of the $K\to\pi\pi$ matrix element
is shown by a straight line with one sigma band.
}
\label{fig:AMP_1_I2}
\end{figure}

\begin{figure}[h]
\includegraphics[width=16.5cm]{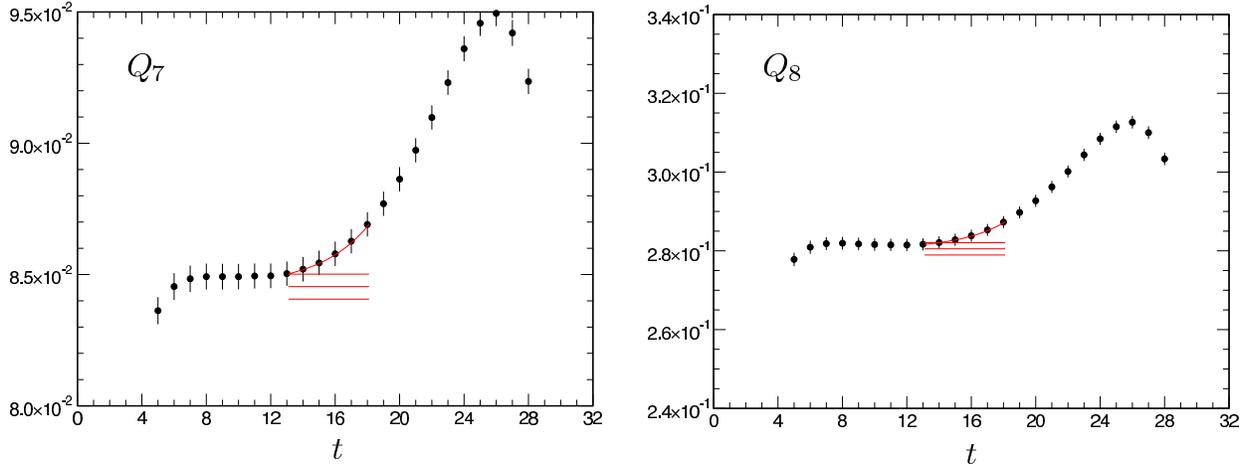}
\caption{
Effective matrix elements
of $M_{i}^{I=2}(t)$ in (\ref{eq:effective_amp})
for the operators $Q_{7,8}$,
following the same convention as in Fig.~\ref{fig:AMP_1_I2}.
}
\label{fig:AMP_78_I2}
%
%------------
\newpage
%------------
\end{figure}

Our results for the effective matrix elements 
for the $\Delta I=3/2$ process 
are plotted in Fig.~\ref{fig:AMP_1_I2} and Fig.~\ref{fig:AMP_78_I2}.
We do not find plateaus.  
Rather they show a large and systematic time dependence 
in the time region $(t_1=0,t_2=4) \ll t \ll (t_K=29)$.
This time dependence can be explained by the effect 
of the two pions propagating in the opposite time directions
({\it i.e.,} around-the-world effect for the two-pion operator) 
as follows.

Including the process 
in which either of the two pions propagates 
in the opposite direction to the other pion, 
the time dependence of the correlation function 
is given by 
\begin{eqnarray}
&&
   G(t,t_1,t_2) 
     = F_1(t) + F_2(t) + F_3(t)
\ ,
\label{eq:G_KPIPI_around} 
\\ 
&& \qquad
    F_1(t) 
  =  \langle 0           |\, W_K(t_K)            \, |       K(p)\rangle \cdot
     \langle K(p)        |\, \overline{Q}(t)     \, |(\pi\pi)(p)\rangle \cr
&& \qquad\qquad\qquad
     \times
     \langle (\pi\pi)(p) |\, W_{\pi\pi}(t_1,t_2) \, |0          \rangle
\cr
&& \qquad\qquad 
  = A' \cdot {\rm e}^{-E_K (t_K-t)} {\rm e}^{-E_{\pi\pi}^{I} (t-t_\pi)} 
\ ,
\label{eq:G_KPIPI_around_1}
\\
&& \qquad 
     F_2(t) 
  =  \langle \pi(0)     | \, W_K(t_K)            \, | K(p)\pi(0) \rangle \cdot
     \langle K(p)\pi(0) | \, \overline{Q}(t)     \, | \pi(p)     \rangle \cr 
&& \qquad\qquad\qquad
     \times
     \langle \pi(p)     | \, W_{\pi\pi}(t_1,t_2) \, | \pi(0)     \rangle 
     \times {\rm e}^{ -m_\pi T }
\cr
&& \qquad\qquad 
  = B' \cdot {\rm e}^{ t_K   ( - E_K ) }
             {\rm e}^{ t_\pi (   E_\pi - m_\pi) }
             {\rm e}^{ t     ( - E_\pi + m_\pi + E_K ) }
             {\rm e}^{ -m_\pi T }
\ ,
\label{eq:G_KPIPI_around_2}
\\
&& \qquad 
     F_3(t) 
  =  \langle \pi(-p)     | \, W_K(t_K)            \, | K(p)\pi(-p) \rangle \cdot
     \langle K(p)\pi(-p) | \, \overline{Q}(t)     \, | \pi(0)      \rangle \cr 
&& \qquad\qquad\qquad
     \times
     \langle \pi(0)     | \, W_{\pi\pi}(t_1,t_2) \, | \pi(-p)     \rangle 
     \times {\rm e}^{ -E_\pi T }
\cr
&& \qquad\qquad
   = C' \cdot {\rm e}^{ t_K   ( - E_K ) } 
              {\rm e}^{ t_\pi ( - E_\pi + m_\pi) }
              {\rm e}^{ t     (   E_\pi - m_\pi + E_K ) }
              {\rm e}^{ -E_\pi T }
\ ,
\label{eq:G_KPIPI_around_3}
\end{eqnarray}
with $t_\pi = (t_1+t_2)/2$,
where the label of the isospin $I$ and $i$ for the operator $Q_i$ are omitted.
States $K(k)$ and $\pi(k)$
are the $K$ meson and the pion with the momentum ${\bf k}=(0,0,k)$
($k=p,0,-p$, ($p=2\pi/L$)),
$E_K=\sqrt{m_K^2+p^2}$ and $E_\pi=\sqrt{m_\pi^2+p^2}$.
The second and third terms in (\ref{eq:G_KPIPI_around}), 
$F_2(t)$ and $F_3(t)$, 
are the contributions from
the around-the-world effect for the two-pion operator.
The constant $A'$ corresponds to the $K\to\pi\pi$ matrix element,
and $B'$ and $C'$ to the $K\pi\to\pi$.
The around-the-world effects for the $K$ meson, 
which are suppressed by ${\rm exp}(- E_K T )$, are neglected
in these equations.
From 
(\ref{eq:G_KPIPI_around_1}),
(\ref{eq:G_KPIPI_around_2}) and 
(\ref{eq:G_KPIPI_around_3}), 
the time dependence of the effective matrix element $M(t)$ 
in (\ref{eq:effective_amp}) can be written as
\begin{equation}
  M(t) = A + B\cdot {\rm e}^{ t S_\pi } + C\cdot {\rm e}^{ - t S_\pi }
\ , 
\label{eq:effective_amp_tdep}
\end{equation}
with $S_\pi = E_{\pi\pi} - E_\pi + m_\pi$,  
where
the constant $A$ is the $K\to\pi\pi$ matrix element,
$M^I_i = \langle K |\, 
           \overline{Q}_i({\bf x},0)\, 
           | \pi\pi; I \rangle$.
The other constants $B$ and $C$ correspond to unphysical matrix elements 
for the $K\pi \to \pi$ process,
including exponential factors for $t_\pi$ and $T$.

We try to extract the $K\to\pi\pi$ matrix elements
by fitting the effective matrix elements $M_i^I(t)$
to the fitting function (\ref{eq:effective_amp_tdep})
with the determined value of $S_\pi$ 
from the correlation function of the pion,
regarding $A$, $B$, and $C$ as unknown parameters.
Results of the fitting are plotted
in Fig.~\ref{fig:AMP_1_I2} and Fig.~\ref{fig:AMP_78_I2}
by curved lines,
and the extracted values of the matrix elements $A=M_i^I$
are shown by straight lines with one sigma band.
We adopt the fitting range $t=[13,18]$.
The chi-squares of the fitting take very small values, 
$\chi^2/n.o.d = 9.6\times10^{-2}, 3.0\times10^{-4}, 4.6\times10^{-4}$
for the $M^{I=2}_{1,7,8}$.
Therefore 
the time dependence of the effective matrix element
is explained well by 
the around-the-world effect for the two-pion operator 
and a value $m_\pi T = 7.55$ in the present work is not large enough 
to neglect this effect.
We also find that
the $K\to\pi\pi$ matrix element can be safely extracted
by the fitting with (\ref{eq:effective_amp_tdep}).
The results of the $M_i^I$ are also tabulated in Table~\ref{table:Result_A2},
where the relations among the matrix elements,
$M_{1}^{I=2} = M_{2}^{I=2} = 2/3 \cdot M_{9}^{I=2} = 2/3 \cdot M_{10}^{I=2}$
due to the Fierz identity, are used.

\begin{figure}[h]
\includegraphics[width=16.5cm]{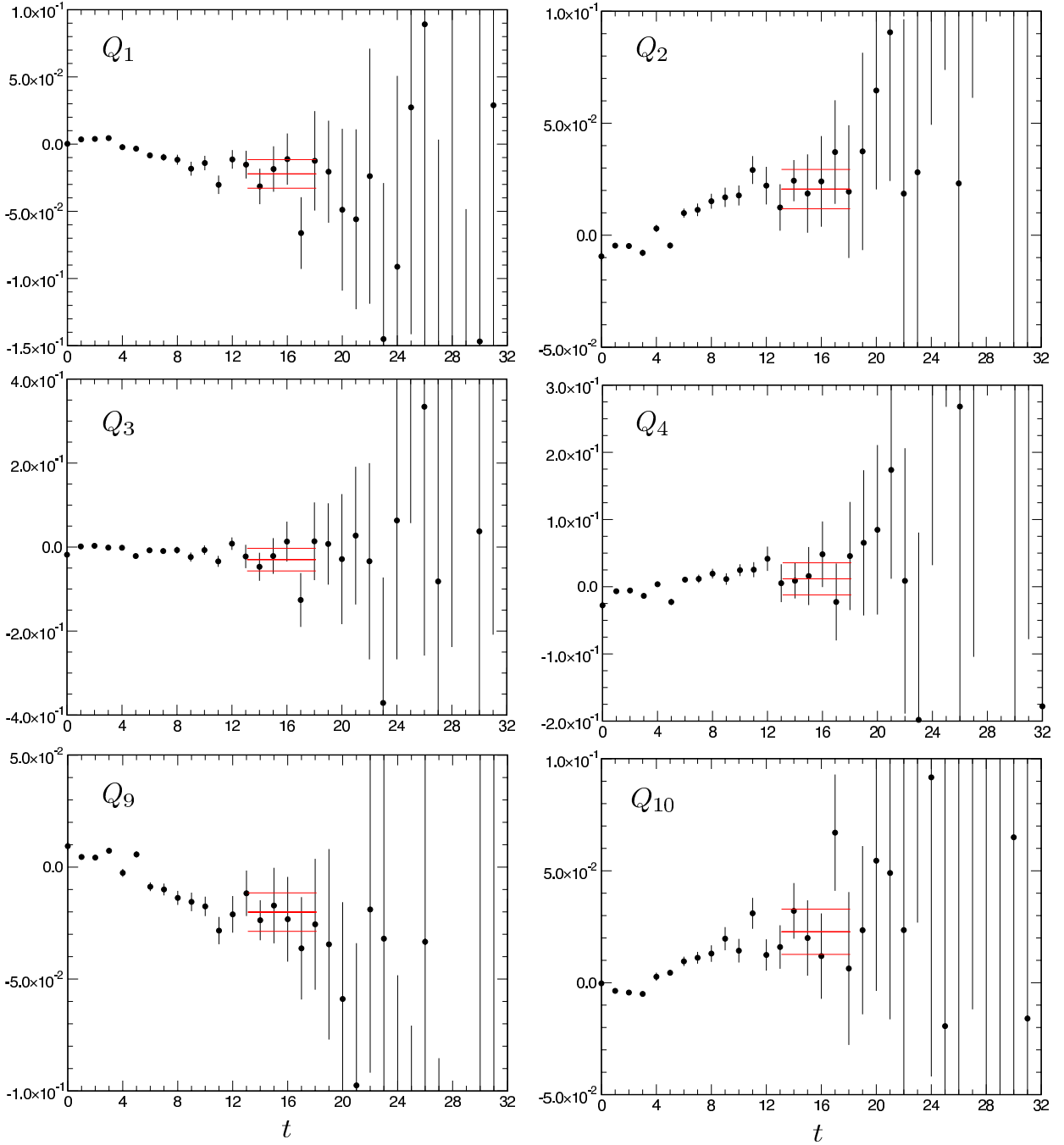}
\caption{
Effective matrix element
of $M_{i}^{I=0}(t)$ in (\ref{eq:effective_amp})
for the LL-type four-fermion operators $Q_{1,2,3,4,9,10}$,
following the same convention as in Fig.~\ref{fig:AMP_1_I2}.
}
\label{fig:AMP_LL_I0}
%
%------------
\newpage
%------------
\end{figure}

\begin{figure}[h]
\includegraphics[width=16.5cm]{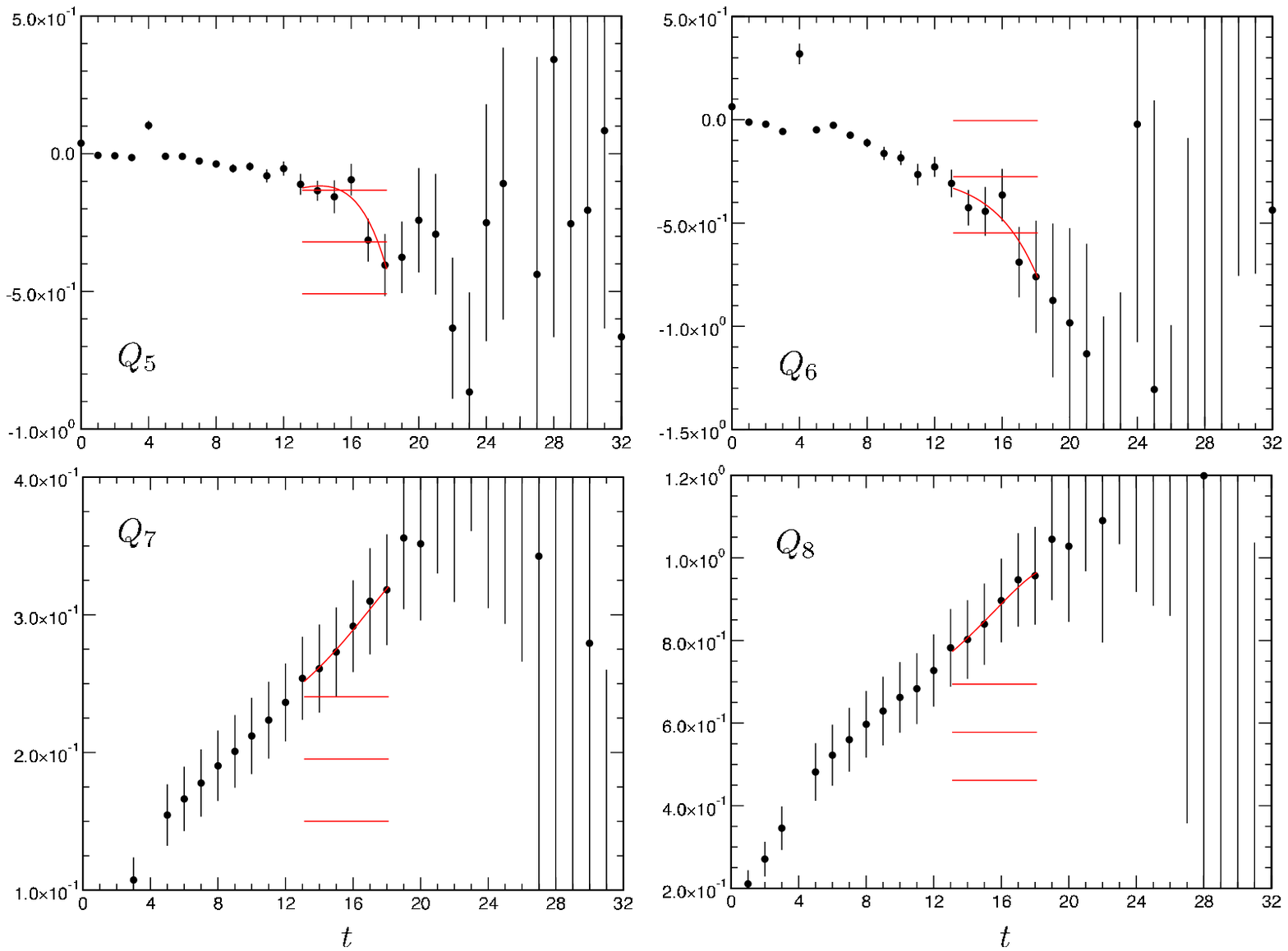}
\caption{
Effective matrix element
of $M_{i}^{I=0}(t)$ in (\ref{eq:effective_amp})
for the LR-type four-fermion operators $Q_{5,6,7,8}$,
following the same convention as in Fig.~\ref{fig:AMP_1_I2}.
}
\label{fig:AMP_LR_I0}
%
%------------
\newpage
%------------
\end{figure}

We should consider this effect also for the $\Delta I=1/2$ process
as for the $\Delta I=3/2$ process.
In Fig.~\ref{fig:AMP_LL_I0}
the effective matrix elements $M_{i}^{I=0}(t)$
for the LL-type four-fermion operators $Q_{1,2,3,4,9,10}$
which have the spin structure 
$\bar{q}_1\gamma_\mu (1-\gamma_5) q_2\, 
 \bar{q}_3\gamma_\mu (1-\gamma_5) q_4$,
are plotted.
We do not observe clear time dependence within the statistical error.
In Fig.~\ref{fig:AMP_LR_I0}
the matrix elements $M_i^{I=0}(t)$ 
for the LR-type four-fermion operators $Q_{5,6,7,8}$
which have the spin structure 
$\bar{q}_1\gamma_\mu (1-\gamma_5) q_2\, 
 \bar{q}_3\gamma_\mu (1+\gamma_5) q_4$,
are plotted.
We see large time dependence for these cases.

Ideally one should extract the matrix elements
by the three parameter fit with (\ref{eq:effective_amp_tdep})
regarding $A$, $B$ and $C$ as unknown constants 
as was done for the $\Delta I=3/2$ process,
to control the around-the-world effect for the two-pion operator.
But the statistics in the present work is not sufficient 
and the time range for the fitting cannot be taken large
to obtain reliable results of the matrix elements.
In the present work,
we therefore extract the matrix elements by a constant fit 
for the LL-type four-fermion operators $Q_{1,2,3,4,9,10}$
assuming that the around-the-world effects for these operators are small, 
given that the results in Fig.~\ref{fig:AMP_LL_I0} 
do not show clear time dependence. 
For the LR-type four-fermion operators $Q_{5,6,7,8}$,
for which we find a large time dependence, 
we use a three parameter fit with (\ref{eq:effective_amp_tdep}).
We adopt the fitting range $t=[13,18]$.
The fits are carried out well and 
the chi-squares of the fitting are small for all cases.
Results of the fitting
are shown in Fig.~\ref{fig:AMP_LL_I0} and Fig.~\ref{fig:AMP_LR_I0},
and tabulated in Table~\ref{table:Result_A0}.
%
%---------------------------------------------------------------------
% @ $K\to \pi\pi$ decay amplitudes
\subsection{ $K\to \pi\pi$ decay amplitudes }
\label{Sec: K to pipi decay amplitudes}
The decay amplitudes can be written 
in terms of the bare matrix element $M_i^I$ obtained on the lattice by
\begin{equation}
   A_I = \langle K | \, H \, | \pi\pi ; I \rangle 
       = \sum_{i=1}^{10} \, M_i^I\, \overline{C}_{i} = \sum_{i=1}^{10} A_I(i)
  \quad , \quad ( \,\, A_I(i) = M_i^I\, \overline{C}_{i} \,\, )
\ ,
\label{eq:A_I_from_M}
\end{equation}
with the effective Hamiltonian $H$ in (\ref{eq:weak_Hamlitonian}),
where
\begin{eqnarray}
&&  \overline{C}_{i} = \sum_{j,k=1}^{10}\, Z_{ij}(q^* a) \, U_{jk}(q^*,\mu)\, C_k(\mu)
\ ,
\label{eq:def_Cbar} 
\\
&&  
   C_i(\mu) = \frac{G_F}{\sqrt{2}}
   \left( V_{us}^* V_{ud} \right)
   \left( z_i(\mu) + \tau y_{i}(\mu) \right)
\ ,
\label{eq:def_C}
\end{eqnarray}
with $\tau = - \left( V_{ts}^* V_{td} \right) 
             / \left( V_{us}^* V_{ud} \right)$.
The functions $C_i(\mu)$ are the coefficient functions 
at the renormalization scale $\mu$ and 
the functions $U_{ij}(q^*,\mu)$ are the running factor
of the operators $Q_i$ from the scale $q^*$ to $\mu$,
which have been given in Ref.~\cite{Review:Buras}.
In the present work
we set $\mu=m_c=1.275\,{\rm GeV}$ 
and adopt the standard model parameters 
tabulated in Table~\ref{table:STD-param}
in estimations of these functions.
The factors $Z_{ij}(q^* a)$ in (\ref{eq:def_Cbar})
are renormalization factors 
which have been calculated
by perturbation theory in one-loop order
in Ref.~\cite{Z-fact:Taniguchi}.
A non-perturbative determination is not yet available.
For the renormalization in the continuum theory,
we adopt
the modified minimal subtraction scheme ($\overline{\rm MS}$)
with naive dimensional regularization scheme (NDR).
We choose two values $q^*=1/a$ and $\pi/a$ as the matching scale
from the lattice to the continuum theory in order to
estimate the systematic error coming
from higher orders of perturbation theory.

Our final results of the decay amplitudes 
at $m_\pi=260\,{\rm MeV}$, $m_K=570\,{\rm MeV}$,
$La=4.4\,{\rm fm}$, and $a=0.091\,{\rm fm}$
are tabulated
with the experimental values
in Table~\ref{table:final_result}.
Here, the direct $CP$ violation parameter $\epsilon'/\epsilon$
is obtained by
\begin{equation}
   {\rm Re}(\epsilon'/\epsilon)
 = \frac{\omega}{\sqrt{2} |\epsilon| }
   \left(
       \frac{ {\rm Im}A_2 }{ {\rm Re}A_2 }
     - \frac{ {\rm Im}A_0 }{ {\rm Re}A_0 }
   \right)
\ ,
\end{equation}
with $\omega = {\rm Re}A_2/{\rm Re}A_0$,
where the experimental value of the indirect $CP$ violation parameter
$|\epsilon| = 2.228\times 10^{-3}$~\cite{STD-param:PDG}
is used in the estimation.
In the table the experimental value of ${\rm Re}(\epsilon'/\epsilon)$ 
is the average value of KTeV, NA48, and E731 measurements~\cite{STD-param:PDG}.
From Table~\ref{table:final_result}
we learn that the dependence on $q^*$
is negligible for most of the decay amplitudes,
but it is very large for ${\rm Im}A_2$.
A nonperturbative determination of the renormalization factor
is necessary to obtain a reliable result for this value.
We find a large enhancement of the $\Delta I=1/2$ process
over that for the $\Delta I=3/2$
at our quark mass $m_\pi = 260\,{\rm MeV}$.
The RBC-UKQCD Collaboration found 
a numerical mechanism for  the enhancement
in Ref.~\cite{DeltaIh_under}.
We confirm this numerical mechanism also in our case.
Our result for $A_0$, particularly for the imaginary part,
still has a large statistical error
so that we do not obtain a nonzero result
for ${\rm Re}(\epsilon'/\epsilon)$ over the error.
Improving statistics by devising a more efficient operator
for the two-pion state
is an important work reserved for the future.

The contributions of the bare matrix element $M_i^{I}$ to
the decay amplitude $A_I$
($A_I(i)$ in (\ref{eq:A_I_from_M}))
are tabulated
in Table~\ref{table:Result_A2} for the $\Delta I=3/2$ and
in Table~\ref{table:Result_A0} for the $\Delta I=1/2$ process.
We find
that the main contribution to ${\rm Re}A_2$ 
comes from the operator $Q_1$ and $Q_2$,
and that to ${\rm Im}A_2$ from $Q_8$.
The main contribution to ${\rm Re}A_0$ comes from the operator $Q_2$
and that to ${\rm Im}A_0$ from $Q_6$.
%
%======================================================================
% @ Conclusion
%
\section{ Conclusions and prospects }
\label{Sec: Conclusions}
We have reported on a calculation of the $K\to\pi\pi$ decay amplitudes
with Wilson fermion action.
In order to realize the physical kinematics,
where the pions in the final state have finite momenta,
we consider the decay process
$K({\bf p}) \to \pi({\bf p}) + \pi({\bf 0})$
in the nonzero momentum frame
with momentum ${\bf p}=(0,0,2\pi/L)$ on the lattice.

We have been able to show a large enhancement of the $\Delta I=1/2$ process
over that for the $\Delta I=3/2$ at our quark mass ($m_\pi=260\,{\rm Mev}$).
However, our result for $A_0$, particularly for the imaginary part,
still has a large statistical error
so that we have not obtained a nonzero result
for ${\rm Re}(\epsilon'/\epsilon)$ over the error.
Improving statistics by devising a more efficient operator
for the $I=0$ two-pion state is necessary
for a precise determination of the decay amplitudes.

Our calculation is carried out away from the physical quark mass.
For calculations near or on the physical quark mass, 
the following extension of the present work is necessary. 
The scattering momentum of the two-pion state
$p=\sqrt{ m_K^2/4 - m_\pi^2}$ at the physical quark mass 
takes a larger value than that in our heavier quark mass case.
Thus, excited states of the two-pion system should be considered
to match the energy with $m_K$.
On our lattice ($a=0.091\,{\rm fm}$ and $La=4.4\,{\rm fm}$)
at the physical quark mass,
the invariant mass of the two-pion state in the moving frame 
with the total momentum ${\bf P}=(1,1,0)$ in units of lattice momentum is 
$400\,{\rm MeV}$ for the ground state
(corresponding to the two-pion state $|\pi(1,1,0),\pi(0,0,0)\rangle$ 
in the noninteracting two-pion case), and 
$490\,{\rm MeV}$ for the first excited state 
($|\pi(1,0,0),\pi(0,1,0)\rangle$).
Therefore we have to calculate the decay amplitude 
for the first excited state.
Calculation of such amplitudes is not trivial, 
because the ground and excited state are mixed by the two-pion interaction.
The decay amplitude for the excited state can be calculated 
by the diagonalization method proposed in Ref.~\cite{Diag:NI}, 
but it has not been applied to the $K$ meson decay.
Therefore, an exploratory study of the method
for the $K$ meson decay is the next step 
in the direction of determining the decay amplitude
at physical quark mass.
%
%===============================================================================
% @ Acknowledgements
%
\section*{Acknowledgments}
This research used computational resources of the K computer
provided by the RIKEN Advanced Institute for Computational Science
through the HPCI System Research Project (Project ID:hp160115).
This work was supported in part 
by Grants-in-Aid of the Ministry of Education, Culture, Sports, Science 
and Technology No.~15H03650, 
and by Multidisciplinary Cooperative Research Program in CCS, 
University of Tsukuba.
%
%==========================================================================
%
%%\newpage

%
%======================================================================
%
\newpage
% \appendix
%
%======================================================================
%  @ Figure
%
%
%=====================================================================
%
%
%=====================================================================
%
%
%=====================================================================
%
%
%=====================================================================
%
%
%
%=====================================================================
%  @ Table
%
\begin{table}[h]
%-------------------
\vspace{-15mm}
%-------------------
\caption{
Lellouch-L\"uscher factor $F^I$ in (\ref{eq:F_LL}).
\hfill
}
\label{table:F_LL}
\begin{tabular}{l r r }
\hline\hline
&  for $I=2$ 
&  for $I=0$  
\\
\hline
$p_c$      &  $0.05993(25)$ 
           &  $0.0512 (50)$
\\
           &  $130.42(54)\,{\rm MeV}$ 
           &  $111(11)   \,{\rm MeV}$ 
\\
$p_c \cdot ( \partial \delta^I(p_c) / \partial p_c )$ 
           &  $-0.0429(78)$
           &  $ 0.24(16)$ 
\\ 
$q_c \cdot ( \partial \phi(q_c) / \partial q_c )$ 
           &  $2.0206(82)$
           &  $1.75(15)$
\\ 
$F^{I}$    &  $51.41(27)$
           &  $63.9(8.6)$
\\
$F^{I} / F|_{\rm free}$ 
           & $0.9701(53)$  
           & $1.21(16)$
\\
\hline\hline
\end{tabular}
\end{table}
%
%%--  1/a = 2.176
%%--  Ch             : I=2   
%%--  sc-mom         :  5.993381D-02  2.48D-04 
%%--           5.993381e-02 * 2.176 * 1000 = 130.41597056
%%--               2.48e-04 * 2.176 * 1000 =   0.539648
%%--  p d(del)/d(p)  : -4.289280D-02  7.80D-03
%%--  q d(phi)/d(q)  :  2.020565D+00  8.22D-03
%%--  LL-factor      :  5.140986D+01  2.70D-01
%%--  Free           :  5.299515D+01  1.14D-01
%%--  LL/Free        :  9.700862D-01  5.25D-03
%%--
%%--  Ch             : I=0   
%%--  sc-mom         :  5.119131D-02  4.96D-03
%%--            5.119131e-02 * 2.176 * 1000 = 111.39229056
%%--                4.96e-03 * 2.176 * 1000 =  10.79296
%%--  p d(del)/d(p)  :  2.416526D-01  1.62D-01
%%--  q d(phi)/d(q)  :  1.750594D+00  1.49D-01
%%--  LL-factor      :  6.389219D+01  8.62D+00
%%--  Free           :  5.299515D+01  1.14D-01
%%--  LL/Free        :  1.205623D+00  1.63D-01
%
%=====================================================================
%
\begin{table}[h]
\caption{
Decay amplitude for the $\Delta I=3/2$ process
at $m_\pi=260\,{\rm MeV}$, $m_K=570\,{\rm MeV}$,
$La=4.4\,{\rm fm}$, and $a=0.091\,{\rm fm}$.
The second column gives
the bare matrix elements $M_i^{I=2}$
for $Q_i$ in the lattice unit.
The other columns are their contribution to
$A_2$ ( $A_2(i)$ in (\ref{eq:A_I_from_M}) )
for $q^*=1/a$ and $\pi/a$.
}
\label{table:Result_A2}
%
%%-- \begin{ruledtabular}
\begin{tabular}{l r rr rr }
\hline\hline
     &
     & \multicolumn{2}{c}{$q^*=  1/a$}
     & \multicolumn{2}{c}{$q^*=\pi/a$}   \\
$ i$ & $M_i^{I}$
     & ${\rm Re}{A_2}\,({\rm GeV})$ & ${\rm Im}{A_2}\,({\rm GeV})$
     & ${\rm Re}{A_2}\,({\rm GeV})$ & ${\rm Im}{A_2}\,({\rm GeV})$ \\
\hline
$ 1$  &$   2.266(18)\times 10^{-3}$ &$-1.890(15)\times10^{-08}$&$ 0                       $&$-1.455(11)\times10^{-08}$&$ 0                       $ \\
$ 2$  &$ = M_{1}^{I=2}            $ &$ 4.335(34)\times10^{-08}$&$ 0                       $&$ 3.926(31)\times10^{-08}$&$ 0                       $ \\
$ 7$  &$   8.454(48)\times 10^{-2}$ &$ 9.182(52)\times10^{-11}$&$ 2.602(15) \times10^{-13}$&$ 2.708(15)\times10^{-10}$&$ 1.970(11) \times10^{-13}$ \\
$ 8$  &$   2.805(16)\times 10^{-1}$ &$-2.444(14)\times10^{-10}$&$-1.5518(87)\times10^{-12}$&$-3.614(20)\times10^{-10}$&$-1.0736(60)\times10^{-12}$ \\
$ 9$  &$ = 3/2 \cdot M_{1}^{I=2}  $ &$-1.281(10)\times10^{-12}$&$ 4.058(32) \times10^{-13}$&$ 3.593(28)\times10^{-12}$&$ 3.678(29) \times10^{-13}$ \\
$10$  &$ = 3/2 \cdot M_{1}^{I=2}  $ &$ 3.940(31)\times10^{-12}$&$-1.896(15) \times10^{-13}$&$ 4.432(35)\times10^{-12}$&$-1.522(12) \times10^{-13}$ \\
\hline                                                                                                                                           
Tot   & \multicolumn{1}{c}{-}       &$ 2.431(19)\times10^{-08}$&$-1.0754(64)\times10^{-12}$&$ 2.463(19)\times10^{-08}$&$-6.611(42) \times10^{-13}$ \\
\hline\hline
\end{tabular}
%%-- \end{ruledtabular}
%
\end{table}
\begin{table}[h]
\caption{
Decay amplitude for the $\Delta I=1/2$ process
at $m_\pi=260\,{\rm MeV}$, $m_K=570\,{\rm MeV}$,
$La=4.4\,{\rm fm}$, and $a=0.091\,{\rm fm}$,
following the same convention as in Table~\ref{table:Result_A2}.
}
\label{table:Result_A0}
%
%%-- \begin{ruledtabular}
\begin{tabular}{l r rr rr }
\hline\hline
     &
     & \multicolumn{2}{c}{$q^*=  1/a$}
     & \multicolumn{2}{c}{$q^*=\pi/a$}   \\
$ i$ & $M_i^{I}$
     & ${\rm Re}{A_0}\,({\rm GeV})$ & ${\rm Im}{A_0}\,({\rm GeV})$
     & ${\rm Re}{A_0}\,({\rm GeV})$ & ${\rm Im}{A_0}\,({\rm GeV})$ \\
\hline
$ 1$&$-2.2(1.1)\times10^{-02}$&$ 1.85(89)\times10^{-07}$&$ 0                     $&$ 1.42(68)\times10^{-07}$&$ 0                     $\\
$ 2$&$ 2.06(88)\times10^{-02}$&$ 3.9(1.7)\times10^{-07}$&$ 0                     $&$ 3.6(1.5)\times10^{-07}$&$ 0                     $\\
$ 3$&$-3.0(2.7)\times10^{-02}$&$ 5.3(4.8)\times10^{-09}$&$ 1.1(1.0)\times10^{-11}$&$ 7.0(6.4)\times10^{-09}$&$ 9.5(8.6)\times10^{-12}$\\
$ 4$&$ 1.2(2.4)\times10^{-02}$&$ 0.6(1.2)\times10^{-08}$&$ 0.8(1.6)\times10^{-11}$&$ 0.7(1.4)\times10^{-08}$&$ 0.7(1.5)\times10^{-11}$\\
$ 5$&$-3.2(1.9)\times10^{-01}$&$ 4.5(2.6)\times10^{-08}$&$ 5.5(3.2)\times10^{-11}$&$ 5.6(3.3)\times10^{-08}$&$ 5.7(3.4)\times10^{-11}$\\
$ 6$&$-2.8(2.7)\times10^{-01}$&$-1.2(1.2)\times10^{-07}$&$-1.6(1.6)\times10^{-10}$&$-1.1(1.1)\times10^{-07}$&$-1.3(1.3)\times10^{-10}$\\
$ 7$&$ 1.95(45)\times10^{-01}$&$ 2.12(49)\times10^{-10}$&$ 6.0(1.4)\times10^{-13}$&$ 6.3(1.5)\times10^{-10}$&$ 4.6(1.1)\times10^{-13}$\\
$ 8$&$ 5.8(1.2)\times10^{-01}$&$-5.0(1.0)\times10^{-10}$&$-3.20(64)\times10^{-12}$&$-7.4(1.5)\times10^{-10}$&$-2.21(45)\times10^{-12}$\\
$ 9$&$-2.01(86)\times10^{-02}$&$ 7.6(3.2)\times10^{-12}$&$-2.4(1.0)\times10^{-12}$&$-2.13(91)\times10^{-11}$&$-2.18(93)\times10^{-12}$\\
$10$&$ 2.3(1.0)\times10^{-02}$&$ 2.6(1.2)\times10^{-11}$&$-1.27(56)\times10^{-12}$&$ 3.0(1.3)\times10^{-11}$&$-1.02(45)\times10^{-12}$\\
\hline                                                                                                                               
Tot   & \multicolumn{1}{c}{-} &$ 5.1(2.8)\times10^{-07}$&$-0.9(1.4)\times10^{-10}$&$ 4.6(2.4)\times10^{-07}$&$-0.6(1.1)\times10^{-10}$\\
\hline\hline
\end{tabular}
%%-- \end{ruledtabular}
%
%------------------------------
\newpage
%------------------------------
\end{table}
\begin{table}[h]
\caption{
Standard model parameters
used to evaluate the decay amplitudes
in the present work (from Ref.~\cite{STD-param:PDG}).
$\tau = - \left( V_{ts}^* V_{td} \right) / \left( V_{us}^* V_{ud} \right)$
and $\Lambda^{(5)}_{\overline{\rm MS}}$ is the lambda QCD for $N_f=5$ theory.
The standard representation of the CKM matrix of Ref.~\cite{STD-param:PDG}
is adopted,
where the $CP$ violation enters entirely through a complex phase of
$V_{td}$, thus $\tau$.
}
\label{table:STD-param}
%
%%-- \begin{ruledtabular}
\begin{tabular}{lr}
\hline\hline
$ m_Z $  &  $  91.1876 \,{\rm GeV} $ \\
$ m_W $  &  $  80.379  \,{\rm GeV} $ \\
$ m_t $  &  $   173.0  \,{\rm GeV} $ \\
$ m_b $  &  $   4.18   \,{\rm GeV} $ \\
$ m_c $  &  $   1.275  \,{\rm GeV} $ \\
$ \Lambda^{(5)}_{\overline{\rm MS}} $ &  $ 231.4 \,{\rm MeV}$ \\
$ \alpha $ ( at $\mu=m_W$ )    &  $ 1 / 129 $ \\
$ \sin^2\theta_W $  &  $  0.230     $ \\
$ G_F            $  &  $  1.166 \times 10^{-5} \,{\rm GeV}^{-2} $ \\
$ V_{ud}         $  &  $  0.97446   $ \\
$ V_{us}         $  &  $  0.22452   $ \\
$ {\rm Re}(\tau) $  &  $  0.0015999 $ \\
$ {\rm Im}(\tau) $  &  $ -0.0006469 $ \\
%%-- $ |\epsilon|     $  &  $ 2.228\times 10^{-3} $ \\
\hline\hline
\end{tabular}
%%-- \end{ruledtabular}
%
\end{table}
%%--
%%-------------------------------------------------------
%%--    
%%--  NEW : PDG2018 :  
%%--    alp_QED = 1.0d0 / 129.0d0   ! alp at Mw
%%--    scw2    = 0.230d0           ! sqr of Winberg angle
%%--  
%%--    mass_W =    80.379d0   ! GeV
%%--    mass_t =   173.0d0
%%--    mass_b =    4.18d0
%%--    mass_c =   1.275d0
%%--  
%%--    mass_Z = 91.1876d0
%%--  
%%--    GF    = 1.16600d-05   ! GeV^-2
%%--    Vud   = 0.97446d0
%%--    Vus   = 0.22452d0
%%--  
%%--    tau_R =  0.0015999d0    ! Standard rep
%%--    tau_I = -0.0006469d0    !   tau = - Vts* Vtd / (Vus* Vud)
%%--  
%%--    eps = 2.228d-03
%%--
%
%======================================================================
%
\begin{table}[h]
\caption{
Results of the $K\to\pi\pi$ decay amplitudes
at $m_\pi=260\,{\rm MeV}$, $m_K=570\,{\rm MeV}$,
$La=4.4\,{\rm fm}$, and $a=0.091\,{\rm fm}$,
and the experimental values~\cite{STD-param:PDG}.
}
\label{table:final_result}
%
%%-- \begin{ruledtabular}
\begin{tabular}{l rrr}
\hline
\hline
%-----------------------------------
& $q^* =1/a$
& $q^* =\pi/a$
& Exp.
\\
\hline
%-----------------------------------
${\rm Re}A_2 \,(\times10^{-8}\,{\rm GeV})$
& $ 2.431 \pm 0.019 $  
& $ 2.463 \pm 0.019 $  
& $ 1.479 \pm 0.004 $
\\
%-----------------------------------
${\rm Re}A_0 \,(\times10^{-8}\,{\rm GeV})$
& $ 51.3  \pm 27.5 $
& $ 45.9  \pm 24.1 $
& $ 33.2  \pm 0.2  $
\\
%-----------------------------------
${\rm Re}A_0 / {\rm Re}A_2$
& $ 21.1  \pm 11.3  $
& $ 18.6  \pm  9.8  $
& $ 22.45 \pm 0.06  $
\\
%-----------------------------------
${\rm Im}A_2 \,(\times10^{-12}\,{\rm GeV})$
& $  -1.0754 \pm  0.0064 $
& $  -0.6611 \pm  0.0042 $
&
\\
%-----------------------------------
${\rm Im}A_0 \,(\times10^{-12}\,{\rm GeV})$
& $ -89.1 \pm 135.5 $
& $ -64.9 \pm 112.4 $
&
\\
%-----------------------------------
${\rm Re}(\epsilon'/\epsilon)(\times 10^{-3})$
& $ 1.94 \pm  5.72 $
& $ 1.95 \pm  5.77 $
& $ 1.66 \pm  0.23 $
\\
%-----------------------------------
\hline
\hline
\end{tabular}
%%-- \end{ruledtabular}
%
\end{table}
%
%%-----------------------------------
%%--  New : PDG2018  
%%--                      q*=1/a                q*=pi/a
%%--    Re(A2)[( -8)GeV] =    2.43070   0.01908     2.46260   0.01929
%%--    Re(A0)[( -8)GeV] =   51.34711  27.46888    45.85188  24.06452
%%--    1/Omega          =   21.12444  11.30310    18.61928   9.77490
%%--    Im(A2)[(-12)GeV] =   -1.07542   0.00641    -0.66113   0.00418
%%--    Im(A0)[(-12)GeV] =  -89.05340 135.51423   -64.88192 112.39492
%%--    EpsP/Eps[(-3)]   =    1.94096   5.72280     1.95436   5.76548
%
%======================================================================
%
\end{document}